\documentclass[10pt,twocolumn]{IEEEtran}
\IEEEoverridecommandlockouts 
\usepackage{graphicx}
\usepackage{verbatim}
\usepackage{amsmath}
\usepackage{enumitem}
\usepackage{bbm}
\usepackage{multirow}
\usepackage{cuted}
\usepackage{lipsum}
\usepackage{widetext}
\usepackage{flushend}

\usepackage{mathrsfs}
\usepackage{color,soul}
\usepackage{algorithm}
\usepackage{algpseudocode}
\usepackage{hyperref}
\usepackage{amsmath,amssymb}
\usepackage{amsmath,amsfonts,amssymb,mathtools}

\usepackage{amsthm}

\usepackage{eurosym}
\usepackage[caption=false]{subfig}

\title{Arbitrage with Power Factor Correction using Energy Storage}

\author{Md Umar Hashmi$^{1}$, Deepjyoti Deka$^{2}$, Ana Bu\v{s}i\'c$^{1}$, Lucas Pereira$^{3}$, and Scott Backhaus$^{2}$
	\thanks{$^{1}$M.U.H. and A.B are with INRIA, DI ENS, Ecole Normale Sup\'erieure, CNRS, PSL Research University, Paris, France.}%
	\thanks{$^{2}$D.D. and S.B. are with Los Alamos National Laboratory, USA }%
	\thanks{$^{3}$L.P. is with Madeira-ITI/ LARSyS and prsma.com, Funchal, Portugal.}%
}

\begin{document}
	
	
	
	\maketitle
	
	\begin{abstract}
The importance of reactive power compensation for power factor (PF) correction will significantly increase with the large-scale integration of distributed generation interfaced via inverters producing only active power. In this work, we focus on co-optimizing energy storage for performing energy arbitrage as well as local power factor correction. The joint optimization problem is non-convex, but can be solved efficiently using a McCormick relaxation along with penalty-based schemes. Using numerical simulations on real data and realistic storage profiles, we show that energy storage can correct PF locally without reducing arbitrage profit. It is observed that active and reactive power control is largely decoupled in nature for performing arbitrage and PF correction (PFC).
Furthermore, we consider a real-time implementation of the problem with uncertain load, renewable and pricing profiles. We develop a model predictive control based storage control policy using auto-regressive forecast for the uncertainty. We observe that PFC is primarily governed by the size of the converter and therefore, look-ahead in time in the online setting does not affect PFC noticeably. However, arbitrage profit are more sensitive to uncertainty for batteries with faster ramp rates compared to slow ramping batteries.
	\end{abstract}
	
\vspace{-8pt}
	\begin{table}[!htbp]
		\caption*{{Nomenclature}} \vspace{-5pt}
		\small
		\label{nomenclature}
	 	\vspace{-4pt}
		\begin{center}
			{\begin{tabular}{p{0.8cm} p{7cm}}
				\hline
			$\eta_{\text{ch}}, \eta_{\text{dis}}$ & Charging and discharging efficiency\\
				$x^i$ & Battery charge change at time $i$\\
				$b^i$ & Battery charge level at time $i$; $b^i= b^{i-1} + x^i$\\
				$P_h^i$ & Active power consumed by inelastic load\\
				$Q_h^i$ & Reactive power consumed by inelastic load\\
				$P_r^i$ & Active power generated by renewable source\\
				$Q_r^i$ & Reactive power output of renewable source\\
				$P^i$ & Active power of inelastic load and renewable generation; $P^i =P_h^i - P_r^i$\\
				$Q^i$ & Reactive power output of inelastic load and renewable generation; $Q^i =Q_h^i - Q_r^i$\\
				$P_B^i$ & Active power output of battery + converter\\
				$P_B^{\max}$ & Maximum active power output of battery + converter\\
				$P_B^{\min}$ & Minimum active power output of battery + converter\\
				$Q_B^i$ & Reactive power output of battery + converter\\
				$S_B^i$ & Instantaneous apparent power output of storage interfaced by converter; $S_B^i = P_B^i + j Q_B^i$\\
				$S_B^{\max}$ & Maximum apparent power output of storage converter\\
				$P_T^i$ & Total active power seen by the grid; $P_T^i = P^i + P_B^i$\\
				$Q_T^i$ & Total reactive power seen by the grid; $Q_T^i = Q^i + Q_B^i$\\
				$p_{\text{elec}}^i$ & Price of electricity at time $i$\\
				\hline
			\end{tabular}}
			\hfill\
		\end{center}
	\end{table}	
\vspace{-12pt}
\section{Introduction}
With the growth of distributed generation (DG) and large-scale renewables, the need to understand their effect on power networks has become crucial. While bulk-renewable generators have well defined rules for performance including that for reactive power, DG owned by small residential consumers has been exempted. This is primarily due to lack of measurement infrastructure and installed DG contributing to a small fraction of total generation. However, in recent years, growing incentives and environmental awareness have resulted in a large number of consumers installing DGs. Policies such as Net-Energy Metering in California has lead to more than 1 million California electricity consumers opting for solar installations by the end of September 2019\footnote{\url{https://www.californiadgstats.ca.gov/}, January, 2019}. Understanding the effects, both operational and financial, of growth in distributed energy resources (DERs) is essential for Distribution System Operators (DSOs) to ensure reliable operation. Since DERs in current markets are not financially rewarded for providing reactive power support, small inverters connected to them primarily output active power and almost no reactive power \cite{hill2012battery}. This is also in compliance with IEEE Standard 1547, which specifies that DG shall not actively regulate the voltage at the point of common coupling \cite{smith2011smart}. As a result, there has been a degradation of the load power factor (PF) \cite{zuercher2012smart}. 

PF denotes the ratio of active power and the apparent power and is measured as $\cos(\phi)$, where $\phi$ denotes the angle between active and reactive power. An alternate definition for PF commonly used in national and ISO level documents is $\tan(\phi)$. As distribution grids are primarily designed to operate close to unity power factor, a systematic degradation in PF can lead to high current, excessive thermal losses, aggravated voltage profiles \cite{kabir2014coordinated}, equipment damage. It has been shown that maintaining a high power factor leads to positive environmental effects due to increased grid efficiency \cite{pfc2007}. To this end, several regional transmission organizations and system operators have operational rules for PF as stated in Table \ref{pfrules}, though primarily for large loads. Note that $|\cos(\phi)|$ implies symmetric rules for leading and lagging power factor.

\begin{table}[!htbp]
		\vspace{-5pt}
		\scriptsize
		\caption {\small{Power Factor Correction Rules}}
		\label{pfrules}
		\vspace{-10pt}
		\begin{center}
			\begin{tabular}{| c | c| }
				\hline
				Utility/Country Name& PF Limit \\
				\hline
				France \cite{francepf} (for $>$ 252 kVA)& $|tan(\phi)| \leq 0.4$ \\
				Portugal \cite{europepf} & $|\cos(\phi)| \geq$ 0.92 \\
				LV consumers Uruguay \cite{uruguay} & $|\cos(\phi)| \geq$  0.92\\
				Germany \cite{germanypf} (for solar users $>$3.68 kVA) & $|\cos(\phi)| \geq$ 0.95 \\
				\hline
				{CAISO}: (a) Wind Generators \cite{pjmpf} & $|\cos(\phi)| \geq$ 0.95\\
				(b) Producers in Dist. Grid \cite{pgetariff},\cite{scetariff},\cite{sdgetariff}& $|\cos(\phi)| \geq$ 0.9 \\
				\hline
				PJM: for Wind Generators\cite{pjmpf} & $|\cos(\phi)| \geq$ 0.95 \\
				ERCOT: for all Generators since 2004 \cite{ercotpf} & $|\cos(\phi)| \geq$ 0.95 \\
				Hydro Ottawa, Canada \cite{ottawa} & $|\cos(\phi)| \geq$ 0.9 \\
				\hline
				\end{tabular}
				\hfill\
				\end{center}
				\vspace{-5pt}	
		\end{table}

\begin{figure}[!htbp]
		\center
		\includegraphics[width=3.5in]{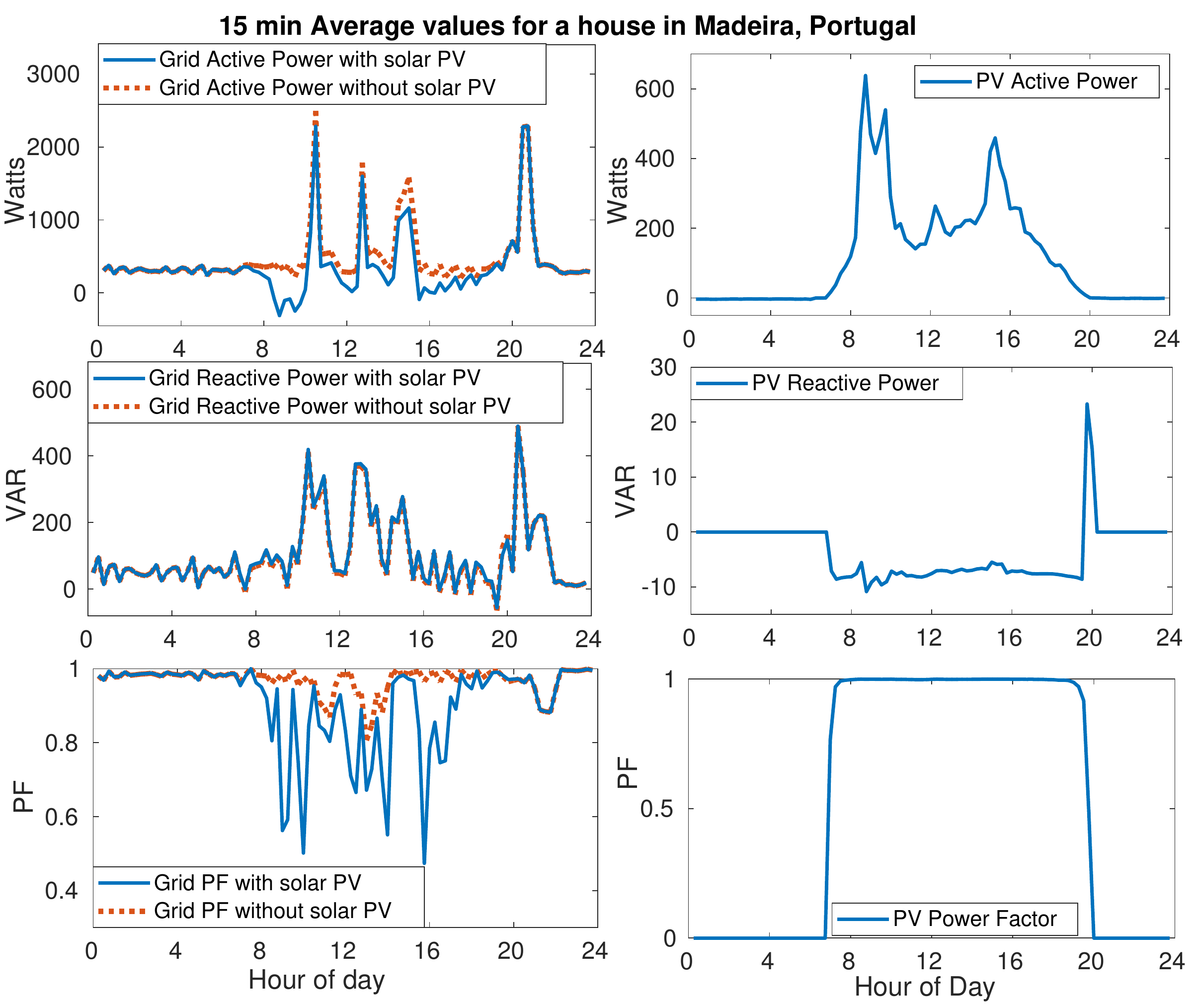}
		\vspace{-5pt}
		\caption{\small{Variation of active, reactive power and absolute value of power factor for PV and the power seen from the grid}}
		\label{madeira}
	\end{figure}

However the PF of residential consumers is also a point of concern. For example, the Smart Islands Energy Systems (SMILE) project, initiated by the European Union in 2017 \cite{smile}, involves data collection at multiple fronts including consumer smart meters in the island of Madeira, Portugal. 
As a case study, 15 minute averaged household consumption and solar generation data on $18^{\text{th}}$ May, 2018 for a representative residential consumer in Madeira is depicted in Fig.~\ref{madeira}.
Note that while PF at night is close to unity, during the day it degrades significantly due to solar output. Thus low load PF may be subjected to norms and penalties \cite{hill2012battery,zuercher2012smart}. Some household smart meters (Eg. Linky smart meters in France) already have reactive power monitoring capability that can implement PF norms \cite{linkyfrance}. {The LV consumers in Uruguay have electricity bills that include penalties for PF degradation \cite{uruguay}. For PF between 0.82 to 0.92, the penalty increases linearly and subsequently becomes a higher rate beyond 0.82.}
\vspace{-5pt}
\vspace{-5pt}	
\subsection{Literature Review}
While additional infrastructure such as capacitor banks \cite{comacond} have been proposed to improve power factor, we focus our work on using conventional energy storage/battery for performing power factor correction, in addition to other functions like arbitrage \cite{ren2016optimal, hashmi2018netmetering}. Note that storage devices generate DC power and hence are connected to the grid through a DC/AC converter/inverter \cite{karagiannopoulos2017hybrid} that are often sized based on the rated active power output capacity. Since such converters output, for majority of time, lower than peak capacity, the remaining capacity can be used for reactive power compensation. Utilizing the storage converter/inverter and power electronics \cite{singh1999review} for power factor correction averts additional investment. The overarching goal of this paper is to demonstrate through novel co-optimization formulations that batteries can be used for PFC without any significant effect on arbitrage profit, for a range of price, consumption and PV settings.
Note that due to the high cost of storage deployment, researchers have proposed using storage for co-optimization additional goals along with energy arbitrage for financial feasibility \cite{kirby2013value}. Inverter reactive power output depends on its control design \cite{ellis2012review}, \cite{turitsyn2010local} and can be governed by terminal voltage and/or active power measurements \cite{karagiannopoulos2017hybrid}, \cite{alam2012distributed}. The authors in \cite{muljadi2004energy} use energy storage for maintaining voltages at wind facilities. Similarly, storage devices have been evaluated using power hardware-in-loop for minimizing losses and voltage fluctuations \cite{taylor2017power}. The authors in \cite{cheng2016co,shi2018using} co-optimize storage for arbitrage, peak shaving and frequency regulation. Unlike the described prior work, we discuss storage for co-optimization of arbitrage and \emph{power factor correction}. Note that contemporary solar inverters in low voltage operate close to unity power factor (UPF) due to no reactive power obligations and hence are practically ineffective for power factor correction.
 \vspace{-5pt}
\subsection{Contribution}
We are interested in using energy storage connected through an inverter for the joint task of arbitrage and PFC. The \textbf{first} contribution of this work is the development of a non-convex mixed-integer formulation to optimize storage for arbitrage and power factor correction in the presence of DG. While the co-optimization problem is non-convex, we demonstrate three different approximation schemes to solve the problem: (a) McCormick relaxation for original non-convex program, (b) receding horizon arbitrage with real-time PFC, and (c) arbitrage with penalty-based PFC. While the McCormick relaxation and real-time PFC policies routinely achieve the optimal solution, the penalty based approach is able to provide best alternatives in scenarios where no feasible solution satisfying PF limits exists. \textbf{Second}, we present a modified penalty-based algorithm that reduces converter usage along with arbitrage and PFC to increase converter lifetime. \textbf{Third}, using realistic pricing, net load (consumption + solar) data and battery parameters, we extensively investigate the achievable ability of storage devices to maintain PF limits without any significant degradation in arbitrage profit. 
\textbf{Fourth}, we consider real-time implementation of our algorithms through the use of Model Predictive Control (MPC) and uncertainty forecasts. We use Auto-Regressive Moving Average (ARMA) processes to model temporally evolving signals in the MPC framework and demonstrate significant benefits from the online algorithms.

The use of dedicated inverters for PV and storage, as analysed in our work, is common when solar is connected on the AC-side. In the case of no storage, the PV inverter can be used for PFC through a relatively simple control algorithm summarized in Appendix~\ref{appendixF}. 
{It is worth mentioning that the inverter control proposed in this work can be used for shared inverters when solar PV is interfaced on the DC-side as the constraints have a similar form. While this paper analyzes PFC with storage incentivized through reactive power charges or utility imposed constraints, the DSO can also involve capacitor banks at the feeder level which are often inexpensive. The selection of utility controlled resources or individual devices will depend on whether costs are socialized or defrayed to the end-user.}

The paper is organized in six sections. Section \ref{sectionii} provides the system description.
Section \ref{sectioniii} formulates the co-optimization problem of performing arbitrage and PFC using storage and discusses multiple solution strategies.
Section \ref{sectioniv} presents an online algorithm using ARMA forecasting and MPC to mitigate the effect of forecast error. Section \ref{sectionv} presents the numerical results. Section \ref{sectionvi} concludes the paper and discusses future directions of research.
\vspace{-5pt}
\section{System Description}
\label{sectionii}
\begin{figure}[!htbp]
		\center
		\includegraphics[width=3.2in]{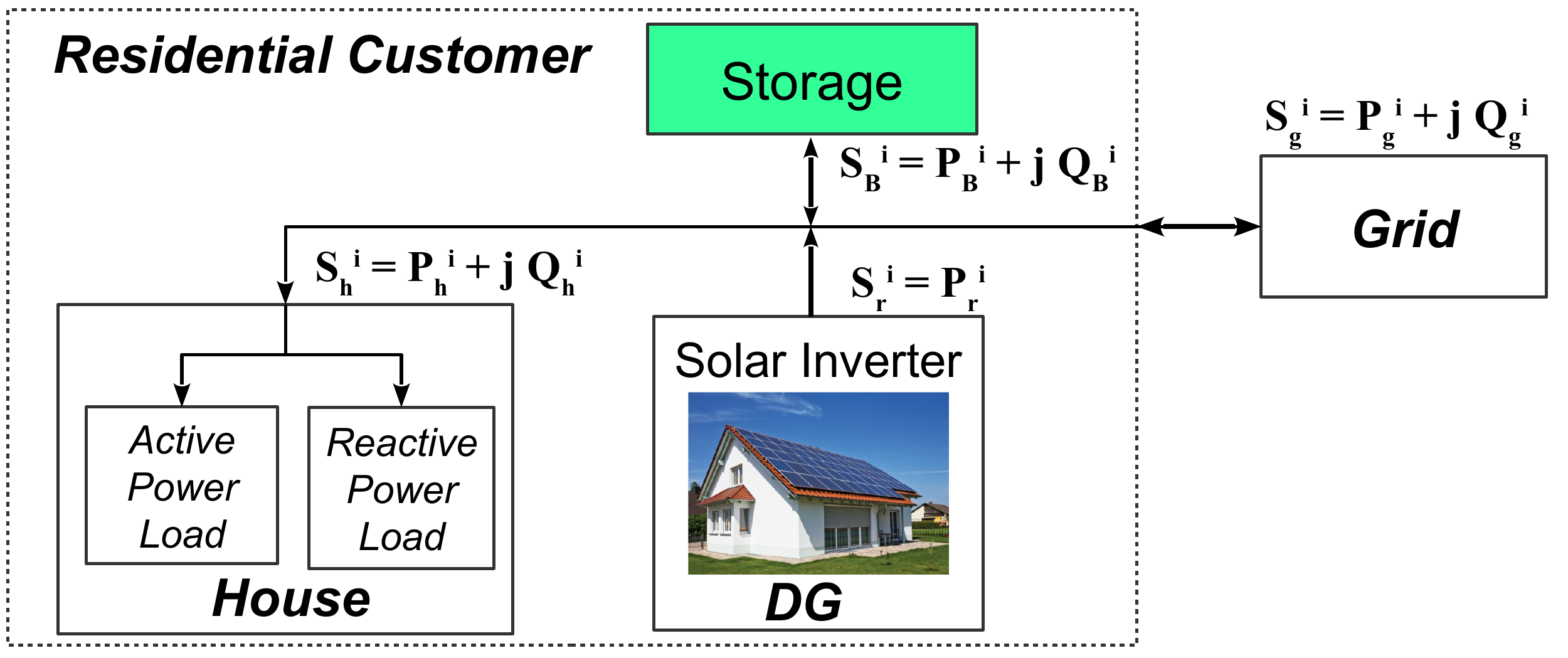} \vspace{-5pt}
		\caption{\small{Residential load block diagram with DG and storage} }\label{block1}
	\vspace{-8pt}
	\end{figure}

The system considered in this work consists of an electricity consumer with inelastic demand, renewable generation (rooftop solar) and energy storage battery. 
The block diagram of the system considered is shown in Fig.~\ref{block1}.
We denote time instant as a superscript of the variable. The apparent power of the load shown in Fig.~\ref{block1}, at $i^{\text{th}}$ time instant is denoted as $S_h^i = P_h^i + j Q_h^i$, where $P_h^i$ and $Q_h^i$ are the active and reactive power consumed. Apparent power of the solar inverter is given as $S_r^i = P_r^i$ where $P_r^i$ is the active power supplied by solar inverter. We assume the solar inverter operates at unity PF. Let us denote the combined load and renewable active and reactive power by $P^i = P_h^i - P_r^i$ and $Q^i = Q_h^i$ respectively. The power factor seen by the grid is the ratio of real power supplied or extracted by the grid over the apparent power seen by the grid. In the absence of storage, the PF before correction is given by 
\vspace{-5pt} 
\begin{equation}
	\text{pf}_{\text{bc}}^i = {P^i}/{\sqrt{{P^i}^2 + {Q^i}^2}}. \vspace{-5pt}
	\label{uncorrect}
	\end{equation}
Observe that $\text{pf}_{\text{bc}}^i$ degrades as $P_r^i$ and $Q_h^i$ increases in magnitude. Next we discuss the battery model considered and its effect on PF. \vspace{-6pt}
	
\textbf{{Battery Model: }}
The storage/battery converter can supply active and reactive power. The apparent power output of energy storage (connected through a converter which is an inverter or a rectifier) is given as $S_B^i = P_B^i + jQ_B^i$, where $P_B^i, Q_B^i$ denote active and reactive power outputs respectively. We consider operation over a total duration $T$, with operations divided into $N$ steps indexed by $\{1,...,N\}$. The duration of each step is denoted as $h$. Hence, $T=hN$. We denote the change in the energy level of the battery at $i^{\text{th}}$ instant by $x^i$; $x^i >0$ implies charging and $x^i <0$ implies discharging. $x^i/h$ denotes the corresponding storage ramp rate with $\delta_{\min} \leq 0$ and $\delta_{\max} \geq 0$ as the minimum and maximum ramp rates (kW) respectively. Let the efficiency of charging and discharging of battery be denoted by $\eta_{\text{ch}}, \eta_{\text{dis}} \in (0,1]$, respectively. The storage active power $P_B^i$ for the $i^{\text{th}}$ instant is related to battery energy as $P_B^i = \frac{[x^i]^+}{h\eta_{\text{ch}}} - \frac{[x^i]^-\eta_{\text{dis}}}{h}$. The active power ramp rate constraint follows as
	\begin{align}
	\footnotesize{
	P_B^i \in [P_B^{\min}, P_B^{\max}]\text{~~with~}P_B^{\min}\text{=}\delta_{\min}\eta_{\text{dis}},~ P_B^{\max}\text{=}\frac{\delta_{\max}}{\eta_{\text{ch}}}},\label{constraintramp}
	\end{align}
	Though the battery charge level is not affected by the reactive power output $Q_B^i$ of the connected inverter, the amount of active power supplied or consumed is dependent upon it due to the line current limitations \cite{kisacikoglu2011reactive}.
	The converter rating is given by the maximum apparent power supplied/consumed, denoted as $S_B^{\max}$ which bounds the instantaneous apparent power $S_B^i$ 
	\begin{equation}
	(S_B^{\max})^2 \geq (S_B^i)^2 = {(P_B^i)^2 + (Q_B^i)^2},
	\label{constraintreactive}
	\end{equation}
Let $b^i$ denote the energy stored in the battery at the $i^{\text{th}}$ step with $b^i = b^{i-1}+ x^i$. To keep the charge in the battery within prescribed limits, the battery capacity constraint is imposed
	\begin{equation}
	b^i \in [b_{\min}, b_{\max}],
	\label{constraintcapacity}
	\end{equation}
where $b_{\min} = SoC_{\min}B_{\text{rated}}$ and $b_{\max} = SoC_{\max}B_{\text{rated}}$. $B_{\text{rated}}$ is the rated capacity and $SoC_{\min}$ and $SoC_{\max}$ are the minimum and maximum level of state of charge respectively.

\textbf{Energy Arbitrage}: The primary use of the storage device considered here is for `Energy arbitrage' which refers to buying electricity when price is low and selling it when price is high, and in effect making a profit. In this work we assume that buying and selling prices of electricity at each instant $i$ are the same and denote it by $p_{\text{elec}}^i$. Under this assumption, the arbitrage profit depends on the varying electricity price but not on the inflexible load. As monetary benefit from arbitrage is \textit{based only on} active power, the operator seeks to minimize the following problem:
	\begin{gather*}
	\text{($P_{arb}$)} \quad
	\min \sum_{i=1}^N {p}_{\text{elec}}^iP_B^i h, \quad \text{subject to,} \text{Eqs.~\ref{constraintramp},~\ref{constraintcapacity}}
	\end{gather*}
We refer the readers to \cite{hashmi2017optimal} for additional details.
\begin{figure}[!htbp]
\center
\includegraphics[width=3.5in]{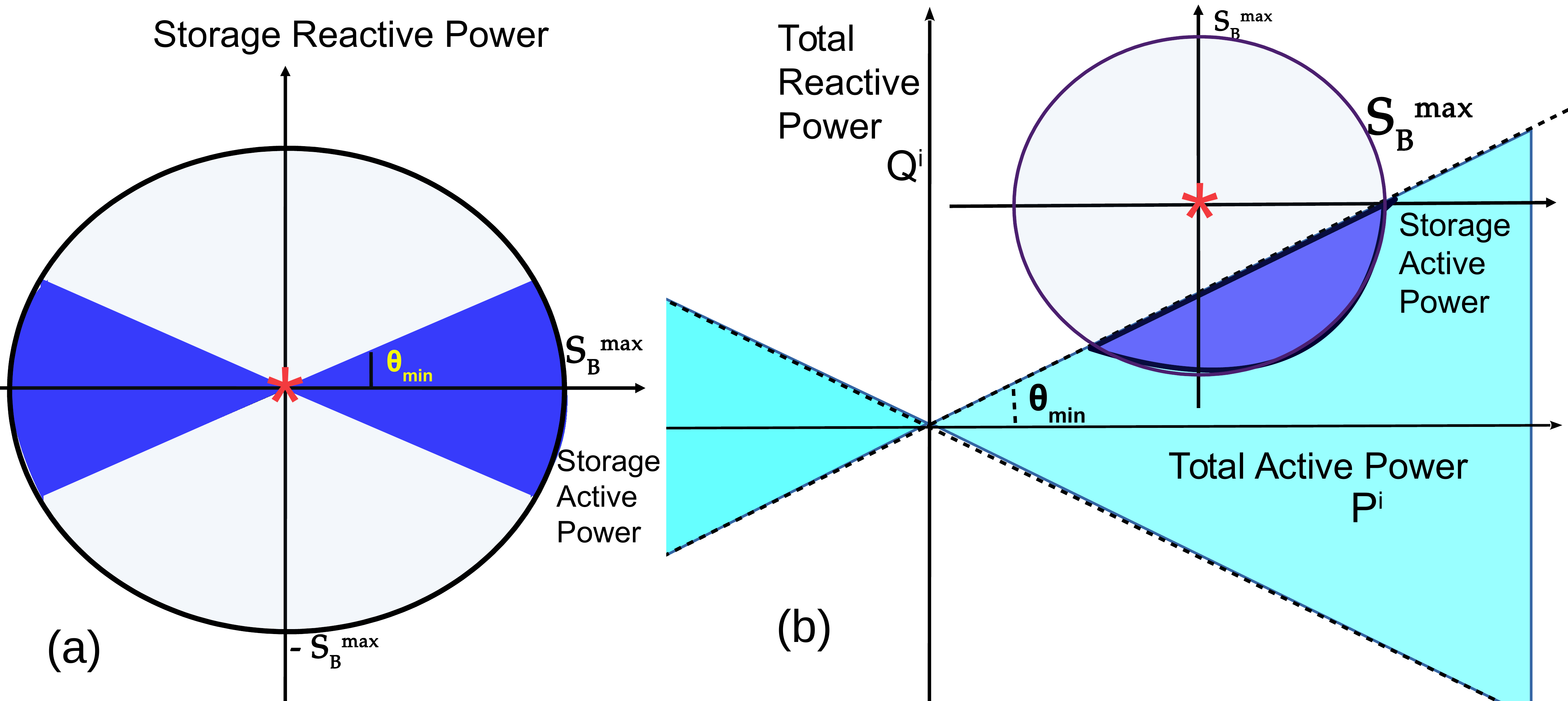} \vspace{-10pt}
\caption{\small{(a) Shows the feasible region of storage in absence of load and DG. The active power out of storage converter, $P_B^i \in [-P_B^{\max},P_B^{\max}]$, and reactive power output ranges as $Q_B^i \in [-\sqrt{(S_B^{\max})^2 -(P_B^i)^2} ,\sqrt{(S_B^{\max})^2 -(P_B^i)^2} ~]$. The utility sets the power factor limit $\text{pf}_{\text{min}}$, which corresponds to power angle $\theta_{\text{min}} = \cos^{-1}(\text{pf}_{\text{min}})$. The deep blue shaded region shows the feasible region of converter operation where output power factor lies within permissible limits of PF. In this plot we assume $P_B^{\max}\geq S_B^{\max}$.
(b) Shows the feasible region in presence of load and DG. The active and reactive power without storage is shown with red asterisk.}}\label{converter}
\end{figure}

\textbf{Power Factor Correction:} Note the power factor formulation in Eq.~\ref{uncorrect}. In the presence of storage, it takes the form
\begin{equation}
	\text{pf}_{\text{c}}^i = {P_T^i}/{\sqrt{(P_T^i)^2 + (Q_T^i)^2}},
	\label{correct2}
	\end{equation}
where total active power and reactive powers are given by
\begin{equation}
P_T^i=P^i + P_B^i, \quad Q_T^i=Q^i + Q_B^i. \label{total_pq}
\end{equation}
It is clear that storage active and reactive power output can either negatively or positively affect the PF seen by the grid. To ensure that the PF is within the permissible limits, the following constraints are imposed
\begin{equation} - k \leq \frac{Q_T^i}{|P_T^i|}  \leq k, ~~ \text{ where } k = \tan (\theta_{\min}).
\label{convexeq}
\end{equation}
We assume that the limits in Eq.~\ref{convexeq} are identical for both leading and lagging PF. Note that the feasible region for the PF constraint as shown in Fig.~\ref{converter}(a) is not convex. In the next section we will formulate a non-convex storage optimization problem and discuss solution strategies.
\vspace{-5pt}
\section{Arbitrage and PFC with Storage}
	\label{sectioniii}
We formulate the co-optimization problem for performing arbitrage and correcting power factor considering active ($P_B^i$) and reactive power ($Q_B^i$) output from storage connected via an inverter. Following the discussion in the preceding section, the objective function is given as
	\vspace{-5pt}
	\begin{gather*}
	\text{($P_{org}$)} \quad
	\min_{P_B^i,Q_B^i} \sum_{i=1}^N {p}_{\text{elec}}^iP_B^i h, \quad
	\text{subject to, } \text{Eqs.~\ref{constraintramp},~\ref{constraintreactive},~\ref{constraintcapacity},~\ref{total_pq},~\ref{convexeq}}.
	\end{gather*}
Eq.~\ref{convexeq} is non-convex but consists of two disjoint convex sets if the active power in the denominator is sign-restricted. 
%
%
{This disjoint nature of PF constraint can be formulated as a mixed-integer convex problem by using McCormick relaxation as described in the first approach in Section~\ref{IIIA}. Section~\ref{IIIB} presents the second sequential approach where arbitrage is optimized first and then PF constraints are corrected for the current instance only. Section~\ref{IIIC} presents a third approach for this problem where we use a convex penalty for PF violations and solve the co-optimization problem by dynamic programming.}
\vspace{-6pt}
\subsection{McCormick Relaxation based approach}
\label{IIIA}
McCormick envelopes are a type of relaxation used in bi-linear programming problems. Using the upper and lower bounds of the bilinear variables, McCormick relaxation approximates the feasible region using a convex quadrilateral \cite{mccormick}.
To use it, we reformulate the non-convex PF constraint Eq.~\ref{convexeq} in ($P_{org}$) as a bi-linear constraint by introducing binary variable $z$ as $|P_T^i| = (2z-1)P_T^i$. 
Let $y = zP_T^i$ denote the bi-linear variable. We then have
 \begin{gather*}
	(2z-1)P_T^i \geq 0 \implies 2y - P_T^i\geq 0.
	\end{gather*}
The McCormick relaxation \cite{mccormick1976computability} for the bi-linear term is represented as follows
	\begin{gather*}
	\vspace{-5pt}
	y \geq z_{lb} P_T^i + P_{lb}^iz -z_{lb}P_{lb}^i,~~
	y \geq z_{ub} P_T^i + P_{ub}^iz -z_{ub}P_{ub}^i,\\
	y \leq z_{lb} P_T^i + P_{ub}^iz -z_{lb}P_{ub}^i,~~
	y \leq z_{ub} P_T^i + P_{lb}^iz -z_{ub}P_{lb}^i.
	\end{gather*}
where $z_{lb}$ ($z_{ub}$) and $P_{lb}^i$ ($P_{lb}^i$) are the lower (upper) bounds for $z$ and $P_T^i$ respectively. As $z_{lb}=0$ and $z_{ub}=1$, the above constraints simplify to
	\begin{gather*}
	y \geq P_{lb}^iz, \quad
	y \geq P_T^i + P_{ub}^iz -P_{ub}^i\\
	y \leq P_{ub}^iz, \quad
	y \leq P_T^i + P_{lb}^iz -P_{lb}^i.
	\end{gather*}
As mentioned in \cite{nagarajan2016tightening}, this McCormick relaxation is exact as one of the variables in the bi-linear term is a binary variable. After simplification, we get the following mixed-integer convex problem ($P_{mr}$) for ($P_{org}$).
	\vspace{-5pt}
	\begin{gather*}
	\text{($P_{mr}$)} \quad
	\min_{P_B,Q_B} \quad \sum_{i=1}^N {p}_{\text{elec}}^iP_B^i h,
	\end{gather*}
	\vspace{-10pt}
	\begin{gather*}
	\text{subject to,} \text{ Eqs.~\ref{constraintramp},~\ref{constraintreactive},~\ref{constraintcapacity},~\ref{total_pq}}\\
	\text{PF constraint: }
	-2ky + kP_T^i - Q_T^i \leq 0,\\
    -2ky + kP_T^i + Q_T^i \leq 0,\\
	\text{Binary variable: } z \in \{0,1\}, ~~
	2y - P_T^i\geq 0,\\
	\text{McCormick constraint:} \quad
	y \geq P_{lb}^iz, \quad y \leq P_{ub}^iz,\\
  y \geq P_T^i + P_{ub}^iz -P_{ub}^i, \quad
	y \leq P_T^i + P_{lb}^iz -P_{lb}^i.
	\end{gather*}
Here $P_{lb}^i = P^i+P_B^{\min}$ is the lower bound of total active power, and $P_{ub}^i = P^i+P_B^{\max}$ is the upper bound. Problem ($P_{mr}$) involving mixed-integer linear constraints can be solved by off the shelf solvers like Gurobi or Mosek that can be called by CVX \cite{cvx}. {Note that ($P_{mr}$) considers arbitrage and PFC at equal footing for all time instances. To study the impact of PFC on arbitrage profit, we propose an approach next where PF of the current instance alone is considered while making optimal arbitrage decisions.}
	\vspace{-5pt}
	\subsection{Receding horizon arbitrage with sequential PFC}
\label{IIIB}
We consider a receding horizon approach ($P_{rh}$) that solves two disjoint optimization problems, denoted as ($P_{sub_1}$) and ($P_{sub_2}$) below, for each time instant $j$ and selects the solution with higher profit and feasibility.
	\vspace{-5pt}
	\begin{gather*}
	\text{($P_{sub_1}$)} \quad
	\min_{P_B,Q_B} \quad \sum_{i=j}^N {p}_{\text{elec}}^iP_B^i h,\\
	\text{subject to,} \text{Eqs.~\ref{constraintramp},~\ref{constraintreactive},~\ref{constraintcapacity},~\ref{total_pq} }\\
	 - kP_T^j \leq Q_T^j \leq kP_T^j, \quad
	 P_T^j \geq 0,
	\end{gather*}
	and the second sub problem is given as
	\vspace{-5pt}
	\begin{gather*}
	\text{($P_{sub_2}$)} \quad
	\min_{P_B,Q_B} \quad \sum_{i=j}^N {p}_{\text{elec}}^iP_B^i h,\\
	\text{subject to, } \text{Eq.~\ref{constraintramp},~\ref{constraintreactive},~\ref{constraintcapacity},~\ref{total_pq}}\\
	 - kP_T^j \geq Q_T^j \geq kP_T^j, \quad
	 P_T^j < 0.
	\end{gather*}
Note that both ($P_{sub_1}$),($P_{sub_2}$) are convex and solve a cumulative arbitrage profit problem, but with {PFC restricted to the current time-instant $j$ only (no look-ahead PFC).} The sub-problems only differ in the sign of the current total active power $P_T^j$. The feasible sub-problem with higher profit sets the storage actions for the current instant $j$. The approach then moves to the next instance $j+1$.

Formulations ($P_{mr}$), ($P_{rh}$) model the PF constraints as hard constraints and ensure their feasibility at every operational point. {However, under some load conditions, PF violations may be unavoidable due to infeasibility with regard to converter and storage constraints. For such cases, we propose an alternate approach next, where we correct PF as best as possible.}
\vspace{-10pt}
	\subsection{Arbitrage with penalty based PFC}
\label{IIIC}
\begin{figure}[!htbp]
		\center
		\includegraphics[width=2.75in]{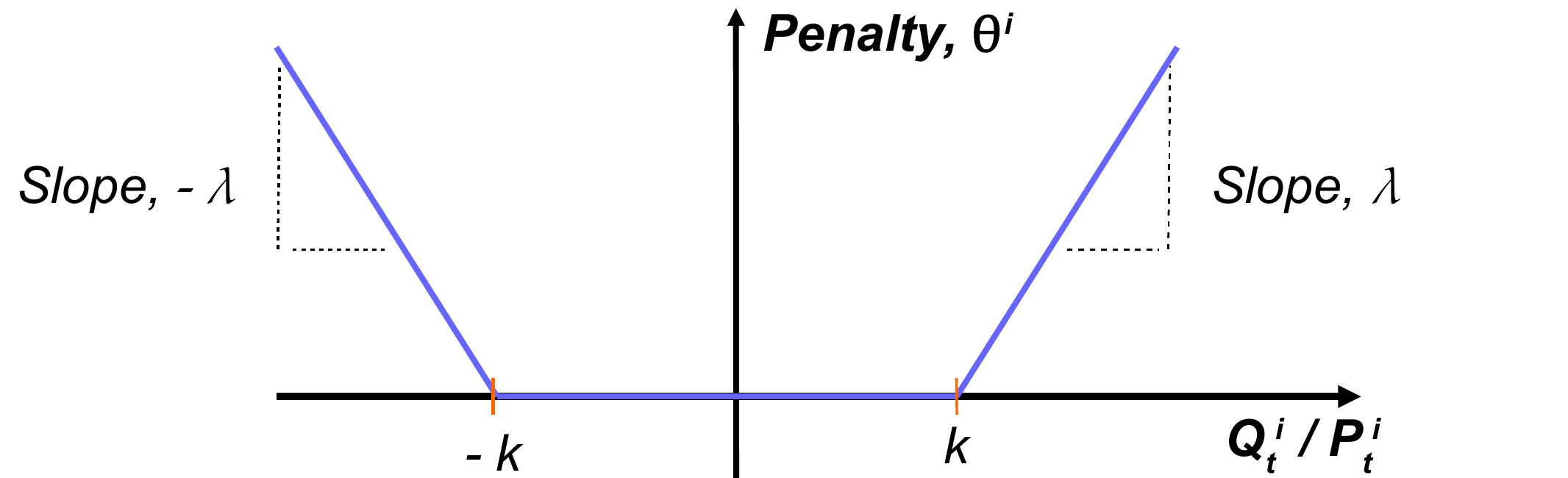}
		\caption{\small{Penalty function with power factor variation} }\label{penaltyfig}
	\end{figure}

	We redefine problem ($P_{org}$) using a penalty function $\mathbb{\theta}^i$ for the power factor. The objective of the new formulation ($P_{plt}$) is given by
	\begin{equation}
\quad \min_{P_B,Q_B} \quad \sum_{i=1}^N \left\{ {p}_{\text{elec}}^iP_B^i h +\mathbb{\theta}^i\right\},
	\end{equation}
	where we define penalty function $\mathbb{\theta}^i$ as
\begin{equation}
 \mathbb{\theta}^i = \lambda \max(0, |Q_T^i| - k |P_T^i| ).
 \label{penaltyconstrainteq}
\end{equation}
Here $\lambda$ represents the constant associated with the linear cost of violating the PF. The structure of the penalty function is shown in Fig.~\ref{penaltyfig}. It is similar to what many utilities impose for industrial consumers (listed in Table~\ref{pfrules}), and to the charge for reactive energy in Uruguay \cite{uruguay} for LV consumers under new consumer contracts proposed in January 2019. For higher values of $\lambda$ with ${p}_{\text{elec}}^iP_B^i h <<  \mathbb{\theta}^i$, PFC will have a higher priority compared to arbitrage and this mode will be similar to Var-Priority Volt/Var Control. For ${p}_{\text{elec}}^iP_B^i h >>  \mathbb{\theta}^i$, the storage converter can be assumed to be operating in Watt-Priority mode \cite{seuss2016analysis}.
The right side of Eq.~\ref{penaltyconstrainteq} will be equal to zero for cases where PF is within permissible limits. The $\max$ term can be modelled as two constraints
	\begin{equation}
 \mathbb{\theta}^i \geq 0,\quad
 \mathbb{\theta}^i \geq \lambda (|Q_T^i| - k |P_T^i| ) \label{penalty1}
	\end{equation}
where absolute value function $|x|$ can further be represented as $(2z-1)x\geq 0$ with binary variable $z \in \{0,1\}$. Eq.~\ref{penalty1} can now be reformulated as
	\begin{equation}
	\begin{split}
	&\mathbb{\theta}^i \geq 0,\quad\mathbb{\theta}^i \geq \lambda (2y_1^i - Q_T^i - 2ky_2^i + k P_T^i).\\
 &2y_1^i - Q_T^i \geq 0, \quad
	2y_2^i - P_T^i \geq 0
	\end{split}
	\label{mccorminckpenalty}
	\end{equation}
Here $y_1^i$ and $y_2^i$ denote bi-linear variables
	\begin{equation}
	y_1^i = z_1^i Q_T^i, \quad
	y_2^i = z_2^i P_T^i
	\label{mccorminckpenaltyx}
	\end{equation}
with binary variables $z_1^i$ and $z_2^i$. As before, we use McCormick relaxation to convert the bi-linear terms in Eq.~\ref{mccorminckpenaltyx} to mixed-integer linear constraints
	\begin{equation}
	\begin{split}
	&y_1^i \geq Q_{lb}^iz_1^i, \quad
	y_1^i \geq Q_T^i + Q_{ub}^iz_1^i -Q_{ub}^i\\
	&y_1^i \leq Q_{ub}^iz_1^i, \quad
	y_1^i \leq Q_T^i + Q_{lb}^iz_1^i -Q_{lb}^i\\
	&y_2^i \geq P_{lb}^iz_2^i, \quad
	y_2^i \geq P_T^i + P_{ub}^iz_2^i -P_{ub}^i\\
	&y_2^i \leq P_{ub}^iz_2^i, \quad
	y_2^i \leq P_T^i + P_{lb}^iz_2^i -P_{lb}^i.
	\end{split}
	\label{mccormicrelaxationpenalty}
	\end{equation}
In these equations, $Q_{lb}^i = Q^i-S_B^{\max}$ and $Q_{ub}^i = Q^i+S_B^{\max}$ denote the lower and upper bounds respectively for total reactive power.

 To summarize, the optimization problem for performing arbitrage and penalized PF violations is given as
\vspace{-5pt}
	\begin{gather*}
	\text{($P_{plt}$)} \quad
	\min_{P_B,Q_B} \quad \sum_{i=1}^N \left\{ {p}_{\text{elec}}^iP_B^i h +\mathbb{\theta}^i\right\}\\
	\text{subject to, } \text{Eq.~\ref{constraintramp},~\ref{constraintreactive},~\ref{constraintcapacity},~\ref{total_pq},~\ref{mccorminckpenalty},~\ref{mccormicrelaxationpenalty}.}
	\end{gather*}

 {The methods described in this section do not consider the degradation of power electronic converters due to usage \cite{yang2010condition}. In the next section, we include a modification of the penalty-based approach with additional cost for minimizing converter usage.}	\vspace{-8pt}
\subsection{Minimizing converter usage with arbitrage and PFC}
\label{IIID}
{It is in the best interest of energy storage owners to minimize the converter operation, measured in apparent power output, to increase their lifetime. In order to emulate this, we add a converter usage component with arbitrage profit and PFC penalty in the objective function of the new optimization problem $(P_{plt}^{conv})$.} 
\vspace{-6pt}
	\begin{equation}
	 \min_{P_B,Q_B} \sum_{i=1}^N \left\{ {p}_{\text{elec}}^iP_B^i h +\mathbb{\theta}^i + \beta \left({(P_B^i)^2 +(Q_B^i)^2}\right)\right\}
	\nonumber
	\end{equation}
	The optimization problem ($P^{conv}_{plt}$) is subject to the same constraints as ($P_{plt}$). 
	{Note that the operational life of a Li-Ion battery, measured using cycle and calendar life, is maximized if the storage operational degradation is matched with the ageing degradation of the battery \cite{hashmi2018limiting}. This can be more accurately added to the co-optimization formulation by including a friction coefficient which eliminates the low returning charge-discharge battery cycles as proposed in \cite{hashmi2018long}. This will be studied in future work.} 
\section{Real-time implementation}
\label{sectioniv}
The previous section discusses multiple approaches to arbitrage maximization and PFC under complete knowledge of future net loads and prices. {In real-world, accurate information of parameters such as consumer load and renewable generation for future time is not known. For real-time implementation, we propose to implement the optimization algorithm in a model predictive framework (MPC) with auto-regressive forecasting for future quantities. In Section~\ref{coOptforecasting}, we describe the forecast model used, and in Section~\ref{coOptMPCorg}, we describe the MPC algorithm.
\subsection{AutoRegressive Forecasting}
\label{coOptforecasting}
We develop a forecast model for future active and reactive power using AutoRegressive Moving Average (ARMA) model, and future electricity prices using AutoRegressive Integrated Moving Average (ARIMA) model. \\
\textbf{ARMA Model for $P$ and $Q$ forecast:}
We define the mean behavior of past values of variable $V$ at time step $i$ as
\begin{equation}
{\bar{V}}^i=\frac{1}{D} \sum_{p=1}^D V_{(i-pN)} \quad \forall i \in \{k,...,N\}, k\geq 1,
\end{equation}
where $N$ is the number of points in a time horizon of 1 day, and $D$ is the number of past days considered. $V$'s forecast is given by: 
\begin{equation}
\hat{V}_i= \bar{V}_i + \hat{M}_i \quad \forall i \in \{k,...,N+k-1\}, k\geq 1,
\label{coOptzhat2}
\end{equation}
where $\hat{M}_i$ is the difference from mean behavior. $\forall i \in \{k,...,N+k-1\}$, $\hat{M}_i$ is modelled as \vspace{-3pt}
\begin{equation}
\hat{M}_k= \sum_{j=1}^J\alpha_j M_{k-j}  + \sum_{u=1}^U \beta_u \delta_k^u, \vspace{-5pt}
\label{xhatk2}
\end{equation}
where $\delta_k^m=(V_{k-mN} - {\bar{V}}_{k-mN})$ and $\alpha_i, \beta_i, \forall i \in\{1,...,U\}$ are constants. 
The weights used in ARMA model, $\alpha_j, \forall j \in \{1,...,J\}, \beta_u, \forall u \in\{1,...,U\}$, are tuned by solving Eq.~\ref{mineqarma2}
\begin{equation}
\min \sum_{i} \{ ||V_i - \hat{V}_i ||^2 + ||\text{\text{norm}}([\alpha^i, \beta^i])||^1 \}. \vspace{-5pt}
\label{mineqarma2}
\end{equation}
We calculate $\hat{P}$ and $\hat{Q}$ using Eq.~\ref{coOptzhat2}. \\
\textbf{ARIMA Model for price $p_{\text{elec}}$ forecast:}
We use ARIMA model of $8^{th}$ lag order and one degree of difference for forecasting electricity price. For variable $X$, the model is denoted as
		\begin{equation}
		\Delta X_{t+1} = \gamma_1 \Delta X_{t} + \gamma_2 \Delta X_{t-1} + ... + \gamma_8 \Delta X_{t-7}
		\end{equation}
		where $\Delta X_{t} = X_{t} - X_{t-1}$.
The coefficients $\gamma_i~ \forall i \in \{1,2,...,8\}$ are tuned based on the historical data, using the \texttt{statsmodels} library for Python \cite{seabold2010statsmodels}.}

\subsection{Model Predictive Control}
\label{coOptMPCorg}
The forecast values are fed to a Model Predictive Control (MPC) scheme \cite{mpcbook} to identity the optimal modes of operation of storage for the current time-instance. Any of the developed schemes from the previous section can be used for the optimization inside MPC. These steps (forecast and MPC) are repeated sequentially and highlighted in online Algorithm $1$: \texttt{ForecastPlusMPC}.


	\begin{algorithm}
		\small{\textbf{Inputs}: {$\eta_{\text{ch}}, \eta_{\text{dis}}, \delta_{\max}, \delta_{\min}, b_{\max}, b_{\min}, S_B^{\max}$}, $b_0$},  {$h, N, T,i=0 $}
		\begin{algorithmic}[1]
			\While{$i < N$}
			\State Increment $i=i+1$,
			\State Forecast $\hat{P}, \hat{Q}$ from time step $i$ to $ N$ using ARMA,
			\State Forecast $\hat{p}_{\text{elec}}$ from time step $i$ to $ N$ using ARIMA,
			\State Co-optimize arbitrage and PFC using inputs $\hat{p}_{\text{elec}}, \hat{P}, \hat{Q}, h$, battery parameters,
			\State Find out battery output: $P_B$ and $Q_B$,
			\State ${b^i}^*= b^{i-1}+[P_B^i ]^+\eta_{\text{ch}} - [P_B^i ]^-/\eta_{\text{dis}}$,
			\State Update $b_0={b^i}^*$,
			\EndWhile
		\end{algorithmic}
		\caption{\texttt{ForecastPlusMPC}}\label{alg:3}
	\end{algorithm}

\vspace{-10pt}	
	\section{Numerical Results}
	\label{sectionv}
	In this section, we demonstrate the performance of our proposed optimization formulations through numerical simulations with real data. We use the following performance indices to measure the performance of different algorithms:
	\begin{enumerate}
		\item \textit{Arbitrage profit}: effectiveness in performing arbitrage
		\item \textit{Power Factor Correction:} is gauged using $3$ indices, using a prescribed PF limit of $0.9$:
			(i) \textit{number of PF violations},
			(ii) \textit{Mean PF}, and (iii) \textit{Minimum PF}.
		\item Converter Usage Factor (CUF): measures usage as
		\begin{equation}
		\text{CUF} = \frac{1}{N} \sum_{i=1}^N {\frac{\sqrt{(P_B^i)^2 + (Q_B^i)^2}}{S_B^{\max}}}
		\end{equation}
		\vspace{-2pt}
	\end{enumerate}	
	The price data for our simulations is taken from NYISO \cite{nyiso} and CAISO \cite{ENOnline}. 
	{For real load and generation profiles, we collect data from the island of Madeira, Portugal. It is worth noting that the island is at a similar latitude as the state of California and has similar patterns of sunshine.} As a benchmark for PFC indices, in Table \ref{resulttab0}, we list the values over a representative day in Madeira, for two nominal cases.
		\begin{table}[!htbp]\label{table2}
		\scriptsize
			\caption {\small{Nominal Cases without Energy Storage for 1 day}}
			\label{resulttab0}
			\vspace{-10pt}
			\begin{center}
				\begin{tabular}{| c | c| c| }
					\hline
					Parameters & no solar, no battery & solar with no battery\\
					\hline
					\hline
					number of PF violations & 8 & 25\\
					mean PF & 0.9735 & 0.9054\\
					min PF & 0.8201 & 0.1587 \\
					\hline
				\end{tabular}
				\hfill\
			\end{center}
		\end{table}
It is evident that with addition of solar, the PF seen by the grid deteriorates with number of PF violations increasing by $200\%$ and minimum PF reached decreasing by $80\%$. We consider different batteries with fixed battery capacity and efficiency but differeing ramp rates and converter capacities. The parameters are listed in Table~\ref{parameters}.
	\begin{table}[!htbp]
	\scriptsize
		\caption {Battery Parameters}
		\label{parameters}
		\vspace{-10pt}
		\begin{center}
			\begin{tabular}{| c | c|}
				\hline
				$B_{\min}$, $B_{\max}$, $B_{0}$ & 200Wh, 2000 Wh, 1000 Wh\\
				\hline
				$\eta_{\text{ch}}=\eta_{\text{dis}}$ & 0.95\\
				\hline
				$\delta_{\max} = - \delta_{\min}$ & 500 W (0.25C-0.25C), \\
				& 2000 W (1C-1C), ~~ 4000 W (2C-2C)\\
				\hline
				$S_B^{\max}$ & $P_B^{\max}$,~~ 0.9$P_B^{\max}$ (undersized), ~~ 1.25$P_B^{\max}$ (oversized)\\
				\hline
			\end{tabular}
			\hfill\
		\end{center}
	\end{table}
{Each ramp rate is described as a ratio of battery capacity over ramp rate. For instance xC-yC ramp rate in Table \ref{parameters} will require 1/x hours to fully charge and 1/y hours to fully discharge. They reflect existing battery ramping. For example, Tesla PowerWall can approximately be denoted as 0.25C-0.25C battery. Faster ramping flywheels can be denoted by a 2C-2C type storage.} By Eq.~\ref{constraintramp}, the ramp rate considered fixes the maximum power $P_B^{\max}$. We define maximum converter capacity $S_B^{\max}$ in terms of $P_B^{\max}$ as listed in the table. The sampling time $h$ is 15 minutes, time horizon $T$ is 24 hours and the power factor limit is 0.9. 
	
	\subsection{Deterministic simulations}
\begin{figure}[!htbp]
		 	\center
		 	\vspace{-10pt}
		 	\includegraphics[width=3.2in]{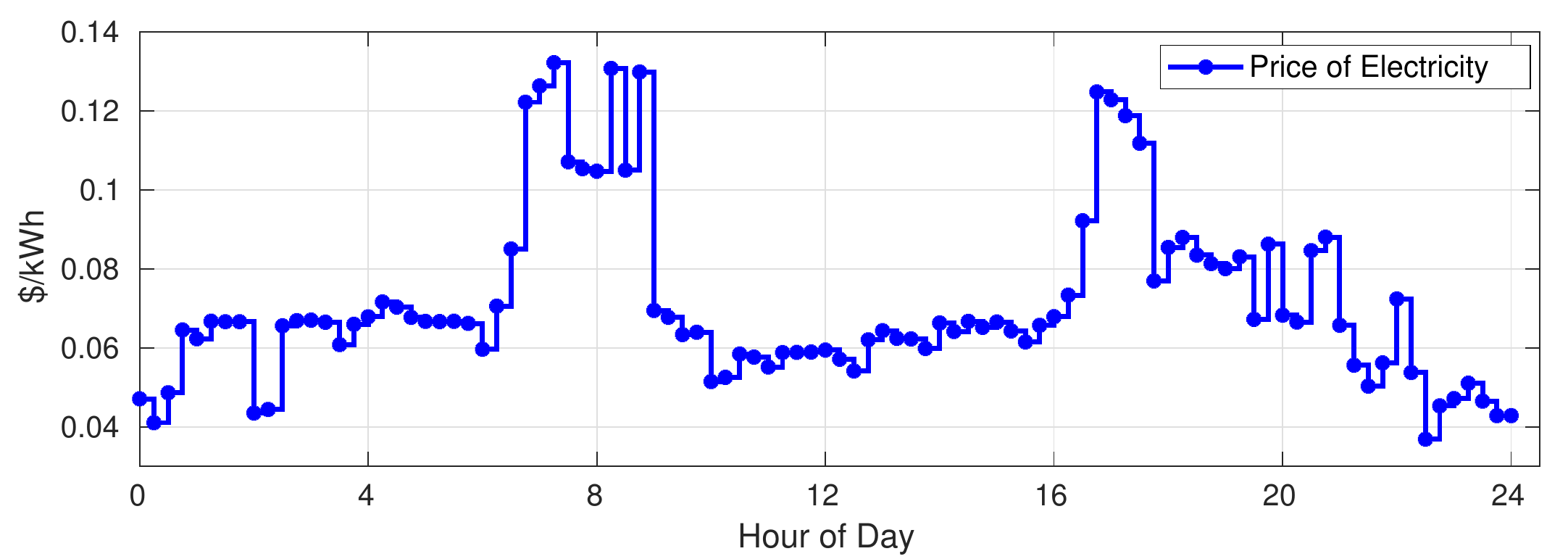} \vspace{-8pt}
		 	\caption{\small{Price data \cite{nyiso} used for 1 day deterministic simulation}}\label{priceSignal}
		 \end{figure}

	{First we discuss results for the five formulations discussed in Section \ref{sectioniii} under knowledge of prices and load data: (a) $P_{arb}$(only arbitrage), (b) $P_{mr}$(McCormick relaxation for arbitrage + PFC), (c) $P_{rh}$(receding horizon arbitrage + sequential PFC), (d) $P_{plt}$(arbitrage + penalized PFC), (e) $P_{plt}^{conv}$(arbitrage + penalized PFC+converter usage). We begin with detailed simulation results for a day and then analyze results over a longer horizon of 2 months.} \\
	{\textbf{Simulations for 1 day:} The price variation for the representative day (96 time instances) is shown in Fig.~\ref{priceSignal}.}
		 
	\begin{table}[!htbp]
	\scriptsize
		\caption {Comparison of arbitrage profit for 1 day}
		\label{resulttab1}
		\vspace{-10pt}
		\begin{center}
			\begin{tabular}{| c| c| c|c| c|c|c|}
				\hline
				$S_B^{\max}=$& Battery & $P_{arb}$ & $P_{mr}$& $P_{rh}$ & $P_{plt}$ &$P_{plt}^{conv}$\\
				\hline
				&0.25C-0.25C 	& \textbf{0.1754} & N.F. & N.F. &0.1747 & 0.1747 \\
				$P_B^{\max}$&1C-1C		& \textbf{0.3367} & 0.3367 & 0.3367 &0.3367 & 0.3367 \\
				&2C-2C		& \textbf{0.4144} & 0.4144 & 0.4144 &0.4144 & 0.4144 \\
				\hline
				&0.25C-0.25C 	& \textbf{0.1728} & N.F. & N.F. &0.1704 & 0.1704 \\
				$0.9P_B^{\max}$&1C-1C		& \textbf{0.3314} & 0.3314 & 0.3314 &0.3314 & 0.3314 \\
				&2C-2C		& \textbf{0.4098} & 0.4098 & 0.4097 &0.4098 & 0.4098 \\
				\hline
				$1.25\times$&0.25C-0.25C 	& \textbf{0.1754} & N.F. & N.F. &0.1753 & 0.1753 \\
				$P_B^{\max}$&1C-1C		& \textbf{0.3367} & 0.3367 & 0.3367 &0.3367 & 0.3367 \\
				&2C-2C		& \textbf{0.4144} & 0.4144 & 0.4144 &0.4144 & 0.4144 \\
				\hline
			\end{tabular}
			\hfill\
		\end{center}
	\end{table}

	\begin{table}[!htbp]
	\scriptsize
		\caption {Comparison of no. of PF violations for 1 day}
		\label{resulttab2}
		\vspace{-10pt}
		\begin{center}
			\begin{tabular}{| c| c| c|c| c|c| c |}
				\hline
				$S_B^{\max}=$ & Battery & $P_{arb}$ & $P_{mr}$& $P_{rh}$ & $P_{plt}$ &$P_{plt}^{conv}$\\
				\hline
				&0.25C-0.25C 	& 27 & N.F. & N.F. &2 & 2 \\
				$P_B^{\max}$&1C-1C		& 26 & 0 & 0 &0 & 0 \\
				&2C-2C		& 24 & 0 & 0 &0 & 0 \\
				\hline
				&0.25C-0.25C 	& 26 & N.F. & N.F. &4 & 4 \\
				$0.9P_B^{\max}$&1C-1C		& 25 & 0 & 0 &0 & 0 \\
				&2C-2C		& 24 & 0 & 0 &0 & 0 \\
				\hline
				$1.25\times$&0.25C-0.25C 	& 26 & N.F. & N.F. &1 & 1 \\
				$P_B^{\max}$&1C-1C		& 26 & 0 & 0 &0 & 0 \\
				&2C-2C		& 25 & 0 & 0 &0 & 0 \\
				\hline
			\end{tabular}
			\hfill\
		\end{center}
	\end{table}

\begin{table}[!htbp]
	\scriptsize
		\caption {Comparison of minimum PF for 1 day}
		\label{resulttab4}
		\vspace{-10pt}
		\begin{center}
			\begin{tabular}{| c| c| c|c| c|c| c|}
				\hline
				$S_B^{\max}=$ & Battery & $P_{arb}$ & $P_{mr}$& $P_{rh}$ & $P_{plt}$ &$P_{plt}^{conv}$\\
				\hline
				&0.25C-0.25C 	& 0.1587 & N.F. & N.F. & 0.8443 & 0.8443 \\
				$P_B^{\max}$&1C-1C		& 0.1587 & 0.9000 & 0.9568 &0.9000 & 0.9000 \\
				&2C-2C		& 0.0545 & 0.9000 & 0.9821 &0.9000 & 0.9000 \\
				\hline
				&0.25C-0.25C 	& 0.1587 & N.F. & N.F. &0.8295 & 0.8295 \\
				$0.9P_B^{\max}$&1C-1C		& 0.1587 & 0.9000 & 0.9000 &0.9000 & 0.9000 \\
				&2C-2C		& 0.0545 & 0.9000 & 0.9000 &0.9000 & 0.9000 \\
				\hline
				$1.25\times$&0.25C-0.25C 	& 0.1587 & N.F. & N.F. &0.8789 & 0.8789 \\
				$P_B^{\max}$&1C-1C		& 0.1587 & 0.9935 & 0.9681 &0.9604 & 0.9000 \\
				&2C-2C		& 0.0545 & 0.9970 & 0.9842 &0.9266 & 0.9000 \\
				\hline
			\end{tabular}
			\hfill\
		\end{center}
	\end{table}
	
We compare the arbitrage profit in Table~\ref{resulttab1} and PF violations Table~\ref{resulttab2} and minimum PF in Table~\ref{resulttab4} respectively for different algorithms and battery settings over the day. Note the arbitrage profit from co-optimizing arbitrage with PFC matches with profit from performing only arbitrage $P_{arb}$, implying performing PFC does not deteriorated energy storage's ability to perform arbitrage. For PF violations, as expected, the number of PF violations for $P_{arb}$ (no PFC) remain close to those in Table \ref{resulttab0}. $P_{mr}$ and $P_{rh}$ are not feasible (denoted as N.F. in results) for battery with slowest ramp rate and small converter as PF violations are unavoidable. However, the other schemes are able to reduce the number of violations drastically. In settings where feasible solution exist, all schemes considered are able to completely avoid any violation.
	\begin{table}[!htbp]
	\scriptsize
		\caption {Comparison of mean PF for 1 day}
		\label{resulttab3}
		\vspace{-10pt}
		\begin{center}
			\begin{tabular}{| c| c| c|c| c|c| c|}
				\hline
				$S_B^{\max}=$ & Battery & $P_{arb}$ & $P_{mr}$& $P_{rh}$ & $P_{plt}$ &$P_{plt}^{conv}$\\
				\hline
				&0.25C-0.25C 	& 0.9062 & N.F. & N.F. & 0.9581 & 0.9562 \\
				$P_B^{\max}$&1C-1C		& 0.9077 & 0.9615 & 0.9938 &0.9426 & 0.9602 \\
				&2C-2C		& 0.8997 & 0.9656 & 0.9983 &0.9378 & 0.9638 \\
				\hline
				&0.25C-0.25C 	& 0.9058 & N.F. & N.F. &0.9512 & 0.9554 \\
				$0.9P_B^{\max}$&1C-1C		& 0.9080 & 0.9610 & 0.9909 &0.9560 & 0.9603 \\
				&2C-2C		& 0.9012 & 0.9659 & 0.9972 &0.9648 & 0.9642 \\
				\hline
				$1.25\times$&0.25C-0.25C 	& 0.9062 & N.F. & N.F. &0.9545 & 0.9567 \\
				$P_B^{\max}$&1C-1C		& 0.9077 & 0.9998 & 0.9962 &0.9742 & 0.9742 \\
				&2C-2C		& 0.8997 & 0.9999 & 0.9987 &0.9478 & 0.9478 \\
				\hline
			\end{tabular}
			\hfill\
		\end{center}
	\end{table}	
	Table~\ref{resulttab3} presents the mean PF. Note that for $P_{mr}$ and $P_{rh}$ in particular, the mean PF for a large converter approaches close to 1, which demonstrates their ability in PFC. However this may lead to overuse of the converter, as evident from CUF listed in Table~\ref{resulttab5}. Here $P_{plt}^{conv}$ provides a way to balance CUF with mean PF as evident from both the tables. Table~\ref{resulttab4} lists the minimum PF measured over the same day. For feasible cases, each algorithm with PFC is able to keep PF equal or above the prescribed limit of $0.9$. However, further analysis would be required to determine penalty functions that motivate or hinder converter usage.

		\begin{table}[!htbp]
		\scriptsize
			\caption {Comparison of CUF for 1 day}
			\label{resulttab5}
			\vspace{-10pt}
			\begin{center}
				\begin{tabular}{| c| c| c|c| c|c| c|}
					\hline
					$S_B^{\max}=$ & Battery& $P_{arb}$ & $P_{mr}$& $P_{rh}$ & $P_{plt}$ &$P_{plt}^{conv}$\\
					\hline
					&0.25C-0.25C 	& 0.7154 & N.F. & N.F. & 0.9198 & \textbf{0.7390} \\
					$P_B^{\max}$&1C-1C		& 0.5520 & 0.5611 & 0.5709 &0.6011 & \textbf{0.5568} \\
					&2C-2C		& 0.4970 & 0.5048 & 0.5059 &0.5248 & \textbf{0.4985} \\
					\hline
					&0.25C-0.25C 	& 0.7693 & N.F. & N.F. &0.9386 & \textbf{0.7823} \\
					$0.9P_B^{\max}$&1C-1C		& 0.5787 & 0.5876 & 0.5989 &0.6183 & \textbf{0.5842} \\
					&2C-2C		& 0.5284 & 0.5375 & 0.5398 &0.5395 & \textbf{0.5302} \\
					\hline
					$1.25\times$&0.25C-0.25C 	& 0.5723 & N.F. & N.F. &0.8045 & \textbf{0.5992} \\
					$P_B^{\max}$&1C-1C		& 0.4416 & 0.4796 & 0.4595 &0.4648 & \textbf{0.4454} \\
					&2C-2C		& 0.3976 & 0.4106 & 0.4058 &0.4208 & \textbf{0.3988} \\
					\hline
				\end{tabular}
				\hfill\
			\end{center}
		\end{table}	
		
\begin{figure}[!htbp]
		 	\center
		 	\vspace{-10pt}
		 	\includegraphics[width=3.5in]{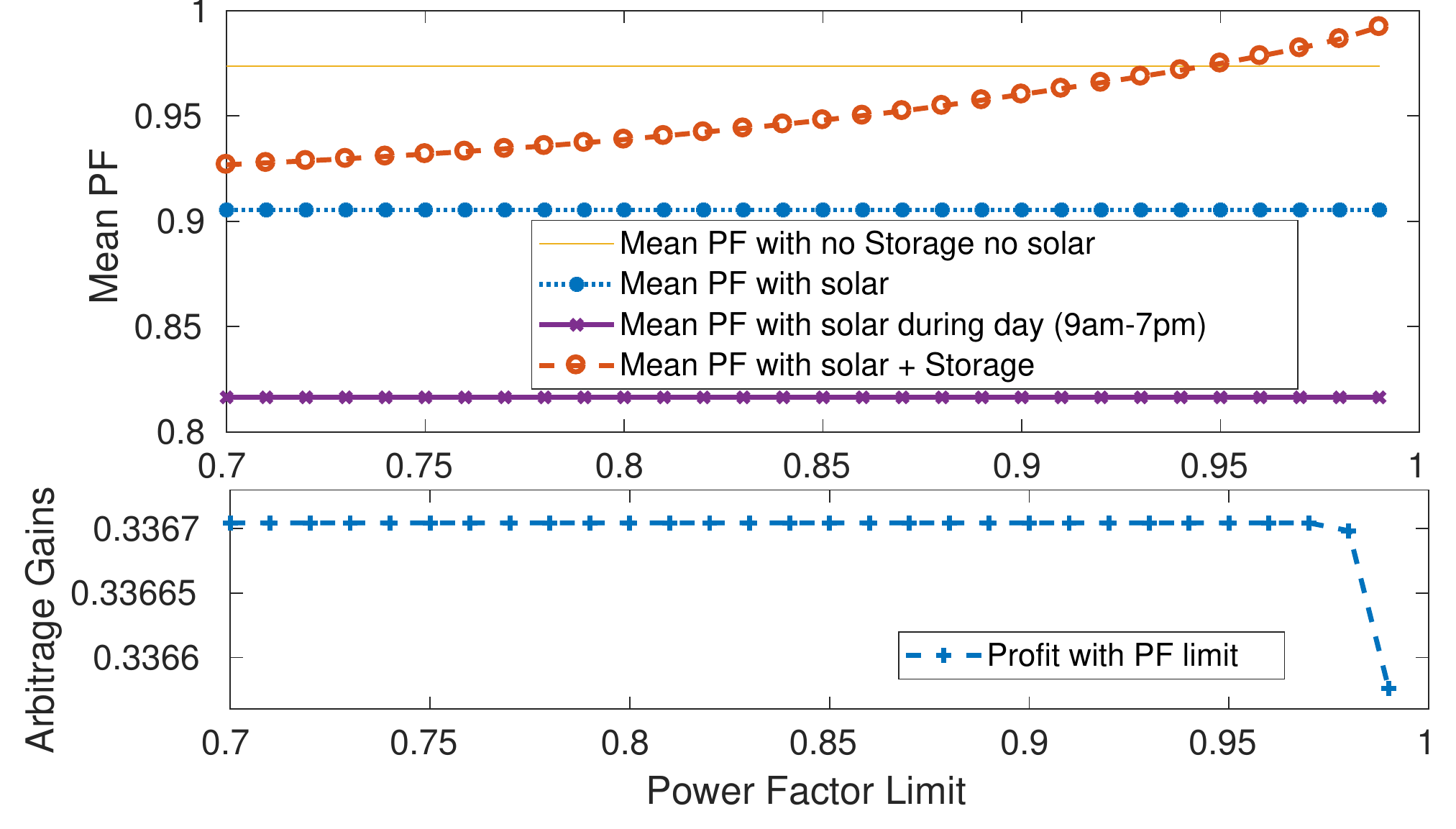} \vspace{-6pt}
		 	\caption{\small{Arbitrage profit with varying PF limit for 1C-1C for $S_B^{\max}$ = $P_B^{\max}$}}\label{pf_limitvary}
		 \end{figure}

It is clear from the mentioned results that storage devices over multiple settings can be used for PFC without any noticeable loss in arbitrage profit.
{Further, from Table~\ref{resulttab4} the PF correction performed myopically in case of $P_{rh}$ coincides with that of look-ahead co-optimizing in $P_{mr}$. This implies that power factor correction does not need look-ahead, unlike arbitrage.} To show the effect of the imposed PF threshold, we consider the 1C-1C battery with converter rating $S_B^{\max} = P_B^{\max}$ in Fig.~\ref{pf_limitvary} and present arbitrage profit, {computed by penalty based co-optimization algorithm $P_{plt}$, over a range of PF thresholds}. Note that the effect of PF limit on arbitrage profits is almost non-existent except for values close to 1.

		
{\textbf{Deterministic simulations for longer time horizon:} We now perform longer time simulation for the months of September and October 2018 in Madeira using a single-day rolling horizon approach. Table~\ref{result2monthsimulation} compares the performance indices for performing only arbitrage ($P_{arb}$) and co-optimizing arbitrage with PFC using ($P_{plt}$). Note that arbitrage profit is maintained despite managing PF for the converter sizes considered.}
\begin{table}[!tbph]
	\caption {Comparison for 2 months of simulation for $P_{arb}$ and $P_{plt}$ for converter $S_B^{\max}= P_B^{\max}$}
	\label{result2monthsimulation}
	\vspace{-10pt}
	\begin{center}
		\begin{tabular}{| c| c| c|c| c|c|}
			\hline
			{Case} & mean & min  & PF   & Profit  & CUF \\ 
			description&PF  &  PF & 	   violations  & (\$)      &\% \\
			\hline
			\multicolumn{6}{|c|}{No Battery} \\
			\hline
			PV  & 0.9544 &0.0048 & 552 &-& -  \\
			\hline
			\multicolumn{6}{|c|}{Solar PV + Battery with $P_{arb}$: only arbitrage}  \\
			\hline
			0.25C-0.25C 	& 0.9554 & 0.0020 & 531 & 10.51  & 75.4 \\
			1C-1C		& 0.9607 & 0.0102
			 & 503  & 30.41 & 63.7  \\
			2C-2C		& 0.9678  & 0.0075 & 385  & 46.65 & 56.1  \\
			\hline
			\multicolumn{6}{|c|}{Solar PV + Battery with $P_{plt}$: arbitrage with PFC}  \\
			\hline
			0.25C-0.25C 	& 0.9715 & 0.1011 & 76 & 10.51  &90.0  \\
			1C-1C		& 0.9553 & 0.2612 & 7 & 30.41 &70.9  \\
			2C-2C		& 0.9483  &0.9000 & 0 & 46.65 & 60.6   \\
			\hline
		\end{tabular}
		\hfill\
	\end{center}
\end{table}
{Next, we discuss results of our storage co-optimization algorithms in the online setting with uncertain knowledge of future electricity price, and active and reactive net load values.}
\subsection{Results with uncertainty}
\begin{figure}[!htbp]
		\center
		\includegraphics[width=3.2in]{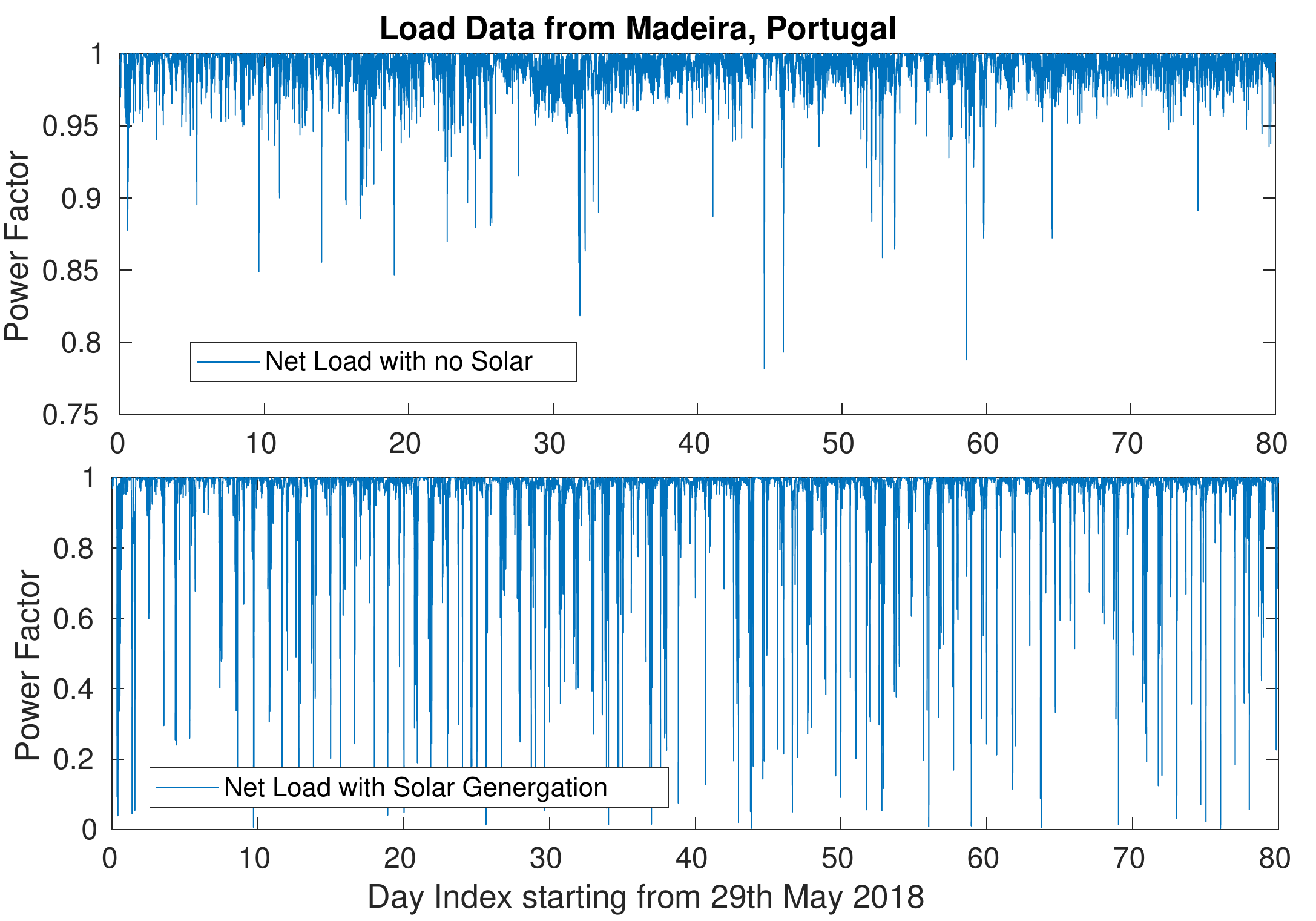}
		\caption{Variation of PF with and without solar PV }\label{loadwithsolar}
	\end{figure}

\begin{figure}[!htbp]
			\center
			\includegraphics[width=3.2in]{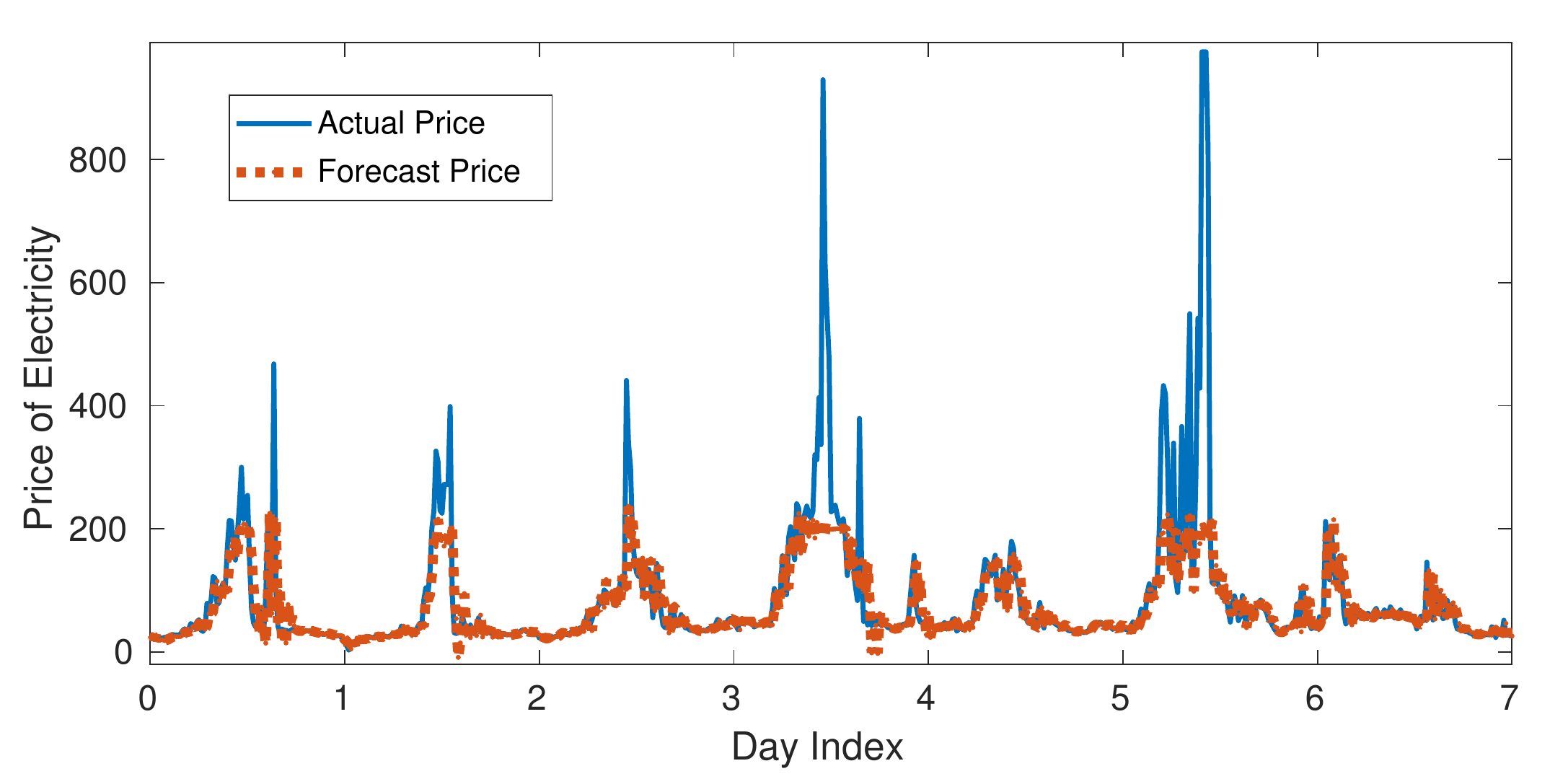}
			\caption{ARIMA Price Forecast}\label{arima}
		\end{figure}

In Section~\ref{sectioniv}, we propose Algorithm $1$ for MPC-based real-time battery control under uncertainty. We implement Algorithm $1$ in rolling horizon with 96 samples (1 day with 15 min sampling) of look-ahead. {We use nine weeks of data starting from 29 May 2018 for training the auto-regressive models for active power, reactive power and electricity price to forecast values for $96$ time-samples. Out-of-sample forecast for the tenth week is then generated for testing. The training data for net load seen by the grid with and without PV is plotted in Fig.~\ref{loadwithsolar}.
Fig.~\ref{loadwithsolar} indicates that inclusion of solar PV has degraded the PF significantly. The performance of the forecast of electricity price signal is plotted in Fig.~\ref{arima}.} Note that the ARIMA model for price misses peaks beyond \$200/MW. However, this drawback of the forecast model is not dominant for batteries with slow ramp rates compared to faster ramping batteries as for such batteries the optimal control action for any price above \$200/MW is to discharge at maximum rate. 
	{In Table~\ref{uncertaintable}, we present average arbitrage profit and PFC indices for the one week of test data using Algorithm $1$ with ($P_{plt}$). 
	{To analyze the effect of uncertainty in forecasts, we compare benefits under Algorithm $1$ with deterministic results for the same period under full knowledge of net load and electricity prices in Table~\ref{deterministictable}.} Note that the arbitrage profit is more sensitive to uncertainty for fast ramping battery. Compared to the deterministic gains, the loss of profits for 0.25C-0.25C battery is only around $3\%$. On the other hand, the loss is close to $35.8\%$ loss for the 2C-2C battery. This is in sync with observations in \cite{yize2018stochastic}. Note in Tables \ref{deterministictable} and \ref{uncertaintable} that PF violations are comparable in the two settings. This consolidates our observations in the previous numerical results that if the converter is not significantly undersized (compared to the maximum active power output of storage), then PFC and subsequently reactive compensation is primarily dependent on the converter size and thus independent of future uncertainties.}
	\begin{table}[!htbp]
	\scriptsize
		\caption {Performance of Algorithm $1$ (real-time implementation with forecast + MPC) for one week}
		\label{uncertaintable}
		\vspace{-10pt}
		\begin{center}
			\begin{tabular}{| c| c| c|c| c|c| c|}
				\hline
				$S_B^{\max}$ & Battery& Profit & Mean& PF &CUF&Min \\
				 & Model& \$ & PF& violations &&PF\\
				\hline
				&0.25C-0.25C 	&2.9996	&0.9704	&13	&0.9075	&0.0488 \\
				$P_B^{\max}$&1C-1C		&4.6840	&0.9465	&0	&0.7032	&0.9000\\
				&2C-2C		&6.0345	&0.9375	&0	&0.6142	&0.9000\\
				\hline
				&0.25C-0.25C 	&2.9718	&0.9652	&25	&0.9324	&0.0656 \\
				$0.9P_B^{\max}$&1C-1C		&4.5934	&0.9684	&4	&0.7258	&0.6268\\
				&2C-2C		&5.9686	&0.9771	&1	&0.6402	&0.5762\\
				\hline
				$1.25\times$&0.25C-0.25C 	&2.9997	&0.9827	&0	&0.7166	&0.9000 \\
				$P_B^{\max}$&1C-1C		&4.6841	&0.9763	&0	&0.5680	&0.9122\\
				&2C-2C		&6.0345	&0.9765	&0	&0.4889	&0.9083\\
				\hline
			\end{tabular}
			\hfill\
		\end{center}
	\end{table}	
	\begin{table}[!htbp]
	\scriptsize
		\caption {Deterministic Performance of $P_{plt}$ for one week}
		\label{deterministictable}
		\vspace{-10pt}
		\begin{center}
			\begin{tabular}{| c| c| c|c| c|c| c|}
				\hline
				$S_B^{\max}$ & Battery& Profit & Mean& PF &CUF&Min \\
				 & Model& \$ & PF& violations & &PF\\
				\hline
				&0.25C-0.25C 	&3.0645&	0.9705&	11&	0.8972&	0.0487 \\
				$P_B^{\max}$&1C-1C		&7.0592	&0.9433	&0	&0.6924	&0.9000\\
				&2C-2C		&9.4113	&0.9364	&0	&0.5868	&0.9000\\
				\hline
				$0.9P_B^{\max}$&0.25C-0.25C 	&3.0385	&0.9644	&26	&0.9278	&0.0883 \\
				&1C-1C		&6.9569	&0.9663	&2	&0.7128	&0.6330\\
				&2C-2C		&9.3096	&0.9754	&1	&0.6149	&0.5688\\
				\hline
				$1.25\times$&0.25C-0.25C 	&3.0647&	0.9831&	0	&0.6985	&0.9033 \\
				$P_B^{\max}$&1C-1C		&7.0593&	0.9764&	0	&0.5495	&0.9000\\
				&2C-2C		&9.4113&	0.9769&	0	&0.4645	&0.9000\\
				\hline
			\end{tabular}
			\hfill\
		\end{center}
	\end{table}
 	\section{Conclusion}
	\label{sectionvi}
In this paper, we propose optimization formulations to operate inverter connected storage devices in distribution grids for co-optimizing arbitrage and power factor correction (PFC), both with or without perfect information. For a majority of cases, we show that the arbitrage profit with PFC converges to the profit achieved when storage performs only arbitrage. The primary reason for PFC being decoupled from arbitrage profit is due to the fact that in most instances, PF can be corrected by adjusting reactive power output. This is primarily governed by converter size and unlike storage active power output, which is constrained by capacity and ramp constraint. We also observe that arbitrage profit of batteries with higher ratio of ramp rate over capacity are more sensitive to uncertainty as they face capacity constraints more frequently.

It is also noteworthy that increasing the converter size would improve the mean PF without any significant change in arbitrage profit for the same ramping battery. In the current work, we consider a stringent case of maintaining PF for every operational point, though the methodology can be extended to the case with penalties on average PF. This work provides multiple avenues for extension. 
{In future work, we will analyze financial incentives and installation costs associated with PFC, and compare PFC at household-level with feeder level control through capacitors. To understand practical applications, our storage control algorithm needs to be extended to the case where the solar and storage share an inverter, as well as combined with load optimization schemes for grid services \cite{mathieu2014arbitraging}. Incorporation of storage lifetime maximization schemes \cite{hashmi2018long} into the optimization formulation is another practical direction of extending our formulation.} Finally, we will research directions to incorporate network power flow constraints pertaining to flow and voltage limits \cite{chance,chance1} into our work on energy storage.
	
	\bibliographystyle{IEEEtran}
	
	\bibliography{pfcor_ref}
	

\appendices
	
\section{Power Factor Correction with Solar Inverter}
\label{appendixF}
{Traditionally solar inverters in LV distribution network operate at close to unity power factor, primarily due to no obligations to supply reactive power. Here we present a special case of the PFC framework presented earlier for storage converter for control of a solar inverter for use in PFC. The system considered here consists of a non-elastic consumer with active and reactive power demand and solar inverter with active power output governed by solar generation, thus an uncontrollable variable. The reactive power output of the solar inverter is controlled so the the PF is corrected as much as possible. The optimization problem for solar inverter for PFC is given as 
\begin{equation}
(P_{\text{solPFC}})\quad \min_{Q_r} \quad \sum_{i=1}^N \mathbb{\theta}^i, \text{subject to, } \text{ Eq.~\ref{constraintreactive}, }
\end{equation}
where we define penalty function $\mathbb{\theta}^i$ as
\begin{equation}
\mathbb{\theta}^i = \lambda \max(0, |Q_h^i + Q_r^i| - k |P_h^i - P_r^i| ).
\label{ch5penaltyeqsolar}
\end{equation}
Note that ($P_{\text{solPFC}}$) has only $Q_r$ as the control variable as $P_h^i - P_r^i$ is known at time $i$. We denote it by $M^i=|P_h^i - P_r^i|$.}

{Since a linear PF penalty implies that the cost of violation is linearly proportional to the amount of violation, therefore in this case, solar inverter can be controlled with no look-ahead (myopically).
The algorithm for solar inverter performing PFC (prioritizing active power) is given in Algorithm \ref{ch5alg:4} \texttt{SolarInverterPFC}.
\begin{algorithm}
	\small{\textbf{Inputs}: {$Q_h^i, P_r^i, P_h^i, S_B^{\max}$}}
	\begin{algorithmic}[1]
		\State Calculate $M^i = |P_h^i - P_r^i|$, Js $= |Q_h^i| - k M^i$
		\State Calculate Slack $= \sqrt{(S_B^{\max})^2 - (P_h^i - P_r^i)^2}$.
		\If{Js $\leq 0$}~~ No PFC required as PF already within limit and set reactive power output of solar inverter $Q_r^i =0$.
		\Else
		\If{$Q_h^i > 0$}~
 $Q_r^i = \max(-Q_h^i + kM^i, -\text{Slack}$).
		\Else~
 $Q_r^i = \min(|-Q_h^i - kM^i|, \text{Slack}$).
		\EndIf
		\EndIf
		\State Return $Q_r^i$.
	\end{algorithmic}
	\caption{\texttt{SolarInverterPFC}}\label{ch5alg:4}
\end{algorithm}
}

\end{document}